\documentclass{egpubl}
\usepackage{envirvis23}
 
\WsPaper           % uncomment for final version of EG Workshop contribution
\usepackage[T1]{fontenc}
\usepackage{dfadobe}  
\usepackage{subcaption}
\usepackage{cite}  % comment out for biblatex with backend=biber
\BibtexOrBiblatex
\electronicVersion
\PrintedOrElectronic
\ifpdf \usepackage[pdftex]{graphicx} \pdfcompresslevel=9
\else \usepackage[dvips]{graphicx} \fi

\usepackage{egweblnk} 
\newcommand{\myparagraph}[1] {\mbox{ }\\ \noindent{\textbf{#1}}}

\title[pyParaOcean: A System for Visual Analysis of Ocean Data]%
      {pyParaOcean: A System for Visual Analysis of Ocean Data}

\author[Jain et al.]
    {\parbox{\textwidth}{\centering 
    Toshit Jain$^1$,
    Varun Singh$^1$,
    Vijay Kumar Boda$^1$,
    Upkar Singh$^1$,
    Ingrid Hotz$^{2,1}$,
    P. N. Vinayachandran$^1$,
    Vijay Natarajan$^1$
    }
    \\
    \parbox{\textwidth}{\centering 
    $^1$Indian Institute of Science Bangalore, India\\
    $^2$Department of Science and Technology (ITN), Link\"oping University, Norrk\"oping, Sweden
    }
    }
\begin{document}

% \linenumbers

% \teaser{
%  \includegraphics[width=\linewidth]{eg_new}
%  \centering
%   \caption{New EG Logo}
% \label{fig:teaser}
% }

\maketitle

\begin{abstract}
Visual analysis is well adopted within the field of oceanography for the analysis of model simulations, detection of different phenomena and events, and tracking of dynamic processes.  With increasing data sizes and the availability of multivariate dynamic data, there is a growing need for scalable and extensible tools for visualization and interactive exploration. We describe pyParaOcean, a visualization system that supports several tasks routinely used in the visual analysis of ocean data. The system is available as a plugin to Paraview and is hence able to leverage its distributed computing capabilities and its rich set of generic analysis and visualization functionalities. pyParaOcean provides modules to support different  visual analysis tasks specific to ocean data, such as eddy identification and salinity movement tracking. These modules are available as Paraview filters and this seamless integration results in a system that is easy to install and use. A case study on the Bay of Bengal illustrates the utility of the system for the study of ocean phenomena and processes. 
%
%\begin{classification} % according to http://www.acm.org/class/1998/
%\CCScat{Computer Graphics}{I.3.3}{Picture/Image Generation}{Line and curve generation}
%\end{classification}
%
\end{abstract}

% \reviewcomments{
% \begin{itemize}
%     \item {needle in figure 2 is not very visible: \todo{same highlight color in all views, thicker needle, extend it}}
%     \item {elaborate more about fig 2 subplots in the caption:\todo{nothing as of yet}}
%     \item {elaborate more on the performance:\todo{add attikan specs}}
%     \item {compare the eddy detection to other methods. eg: FFH21(that uses ssh and velocity both): \todo{add a statement saying it uses general techiniques and can be extended to other techniques}}
%     \item {elaborate more about the type of grids pyparaocean works on - regular/rectilinear/irregular: \todo{architecture description should mention pyparaocean works with ocean data which is usually netCDF and uses rectilinear/regular grids}}
%     \item {availability of pyparaocean:\todo{add a concluding statement saying that we plan to release it in the future}}
%     \item {need to put more emphasis on the 3D part?\todo{add to the eddy tool saying it's 3D}}
%     \item {minor grammatical errors: \todo{ Need to check with grammarly once}}
%     \item {can update upkar's figures to match the other aesthetics:\todo{remove the map, remove the ocean, add a grey solid slice for those things}}
%     \item{remove line numbers}
% \end{itemize}
% }

%-------------------------------------------------------------------------
\section{Introduction}\label{sec:introduction}
Understanding ocean data is paramount to predicting extreme weather events such as hurricanes and tsunamis, better understanding of planetary scale processes like global warming, and sustainably managing and preserving ocean resources and its marine life. Visualizing ocean data is challenging due to the presence of multiple fields and parameters that change with time. Ocean currents are undeniably the biggest factor that maintain the heat balance of the ocean-atmosphere system and affect the movement of minerals and salt.  Mesoscale eddies, that span from tens to hundreds of kilometers in diameter and have a lifespan that can range from days to months~\cite{robinson2010mesoscale}, are ubiquitous in the oceans. They play a big role in transporting heat and mass within the oceans~\cite{mcwilliams2008nature}. They also have a big impact on the ecology of the ocean and on the biogeochemical processes~\cite{mcneil1999new,benitez2007mesoscale}.

With the strides in collection and generation of ocean data~\cite{fraser2006data,rosenblum1989visualizing}, there is a need for tools that support effective visualization of the data, and are scalable to keep pace with the ever expanding resolutions and sizes of ocean datasets.

\subsection{Related work}
%\begin{itemize}
%    \item Oceanography visualization frameworks such as RedSeaAtlas, VAPOR, Ferret
%    \item Individual visualization tools from the survey by Xie et al. 
%    \item Eddy detection and visualization tools
%    \item Research gaps and the need for pyParaOcean
%\end{itemize}
%
Visualization in oceanography is a challenging area of research due to the rapidly increasing size of data, the heterogeneity and multivariate nature of the data, and the inherent complexity of the ocean phenomena. The use of general purpose analysis and visualization software such as Matlab, Tecplot, AVS, and Paraview is prevalent in the community. However, oceanographers often use tools developed specifically for ocean data, such as Ferret~\cite{ferret}, pyFerret~\cite{pyferret}, and Copernicus MyOcean~\cite{myocean}. These specialized tools provide multiple functionalities and produce 2D views of the data. 

A few software frameworks developed within the visualization community provide 2D and 3D data visualization capabilities.  COVE~\cite{grochow2008cove} is a collaborative ocean visualization environment that supports interactive analysis of ocean models over the web. RedSeaAtlas~\cite{afzal2019redseaatlas}  supports the selection of regions in a 2D map and provides exploratory views of winds, waves, tides, chlorophyll, etc. over the Red Sea. OceanPaths~\cite{nobreocean2015} is a multivariate data visualization tool  that computes pathways tracing ocean currents and supports the plotting of different high-dimensional data along the pathways. This enables the study of correlations between different oceanographic features. 

An oceanographer's analysis workflow includes a few common tasks~\cite{grochow2008cove} such as inspection of temperature and salinity distributions and vertical cross sections, compare recently measured salinity data against model data, inspect and analyze current vorticities and circulation based on flow data, and analyze extreme events. Isosurfaces and volume rendering are natural choices for visualization of 3D temperature and salinity distributions~\cite{dinesha2012uncertainty,park2004visualization}. However, visualization of the dynamically changing distributions is a challenge.  VAPOR~\cite{li2019vapor} is one of the few tools that provides efficient 3D visualization for oceanography and atmospheric science applications. The VAPOR data collection (VDC) data model supports interactive visual analysis of large data sizes on modern GPUs and commodity hardware.  

Xie et al.~\cite{xie2019survey} and Afzal et al.~\cite{afzal2019state} present surveys of visual analysis methods and tools developed for ocean data. Xie et al. classify the visual analysis tasks into four categories, namely study of different environmental variables, ocean phenomena identification and tracking, discovery of patterns and correlations, and visualization of ensembles and uncertainty.  Further, they identify different opportunities and unexplored areas for future work including efficient and scalable methods for data processing and management, identification of features at multiple scales,  and immersive platforms. While we recognize the availability of several methods for oceanography visualization, we note that they are often standalone solutions. We aim to leverage the extensive advancement in visualization technology as implemented in Paraview while providing functionalities and views that are specific to ocean data. 

\subsection{Contributions}
We present pyParaOcean, a system for interactive exploration and visual analysis of ocean data. The system leverages the power of Paraview~\cite{AhrensGeveciParaview2005} to enable scalable visualizations of data available from ocean models while supporting a multitude of tasks and functionalities that are specialized for oceanography. The visualization capabilities of pyParaOcean are available via a seamless integration into Paraview using plugins. Key features of the system include
\begin{itemize}
\item 3D field visualization to study salinity and temperature distribution with support to display and explore dynamically evolving isovolumes. 
\item Ocean current visualization with a support for different seeding strategies for streamlines and pathlines.
\item Vertical section views and parallel coordinates plot that support interactive selection and slicing of the data.
\item Identification and tracking of salinity movement via extraction of surface fronts.
\item Visualization and tracking of eddy features.
\item An extensible design that supports incorporation of new functionalities as \textit{filters} in Paraview.
\end{itemize}
We present the results of an exploration of the Bay of Bengal, performed in collaboration with an oceanographer, as a case study to demonstrate the utility of the system.

\section{Ocean data}\label{sec:data}
Oceanographers typically deal with large multivariate spatio-temporal datasets -- time-varying scalar or vector fields over a three dimensional region. The data is generated using simulations, satellite imagery, sensors on buoys, or in-situ physical observations. With strides in high performance computing, higher resolution sampling, and the increasing number of observables, the size of such datasets is rapidly increasing. Reanalysis datasets combine a numerical simulation model with observational inputs to furnish data that is spatio-temporally consistent.
Ocean data contains strong temporal and spatial processes involving complex interactions between multi-scale entities~\cite{xie2019survey}. It is analyzed on a variety of scales, from small-scale features such as eddies and fronts, to large-scale features such as ocean basins and circulation patterns.

All visualizations in this paper are generated using two datasets, the Red Sea and Bay of Bengal.

\myparagraph{Red Sea}: This dataset~\cite{Toy17} was made available as a part of the IEEE SciVis 2020 contest. It is a 50 member ensemble of three-dimensional scalar and velocity fields. The data is regularly sampled on a 500 $\times$ 500 $\times$ 50 grid over 60 time steps covering a whole month of simulation time. Ensembles are the outputs of the simulated models with different parameters and initial conditions, and they may vary significantly even with a small change in parameter values. The members are the forecasts from MITgcm setups configured for the 30$^{\circ}$E - 50$^{\circ}$E and 10$^{\circ}$N -  30$^{\circ}$N domain that spans the entire Red Sea. They are implemented in Cartesian coordinates with a horizontal resolution of 0.04$^{\circ}$ $\times$ 0.04$^{\circ}$ (4~km) and 50 vertical layers, with a surface spacing of 4~m and a bottom spacing of 300~m. The dataset is available in the NetCDF format.

\myparagraph{Bay of Bengal}: This dataset is generated by a reanalysis product and available from the Nucleus for European Modelling of the Ocean (NEMO) repository~\cite{Madec2008}, with a daily resolution spanning the months of July-August 2020, a total of 62 time steps. The data is available in NetCDF format, with a 1/12$^{\circ}$ latitude-longitude resolution. Salinity measurements are available at 50 vertical levels, ranging from 1~m resolution near the surface to 450~m resolution towards the sea floor, including 22 samples in the upper 100~m. The Bay of Bengal, a geographical region confined by longitudes 75$^{\circ}$E and 96$^{\circ}$E and latitudes 5$^{\circ}$S to 30$^{\circ}$N, with depth up to 200~m, is extracted from this data.

\section{pyParaOcean: Architecture}\label{sec:pyparaocean-architecture}
\begin{figure}
  \centering
  \includegraphics[width=\linewidth]{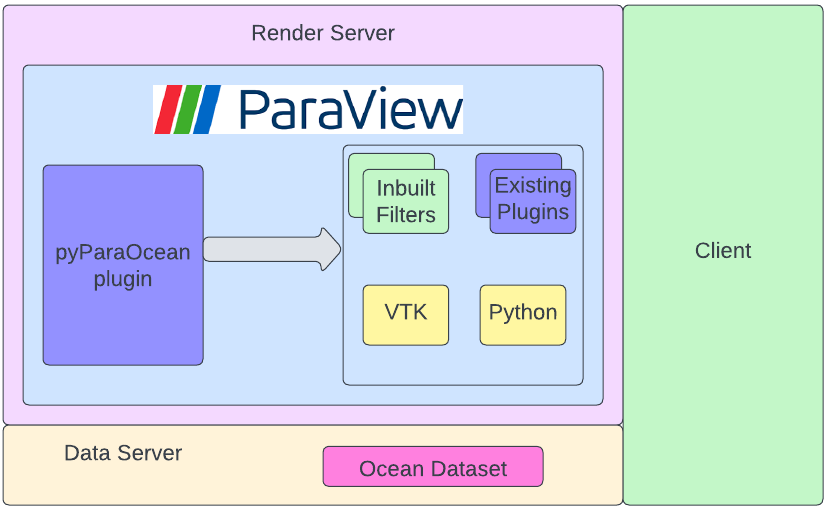}
  \caption{pyParaOcean system architecture. The plugin includes multiple specialized filters for visualizing ocean data that seamlessly integrate with the high performance capabilities of Paraview.}
  \label{fig:architecture}
\end{figure}
\begin{figure*}
    \centering
    \includegraphics[width=\linewidth]{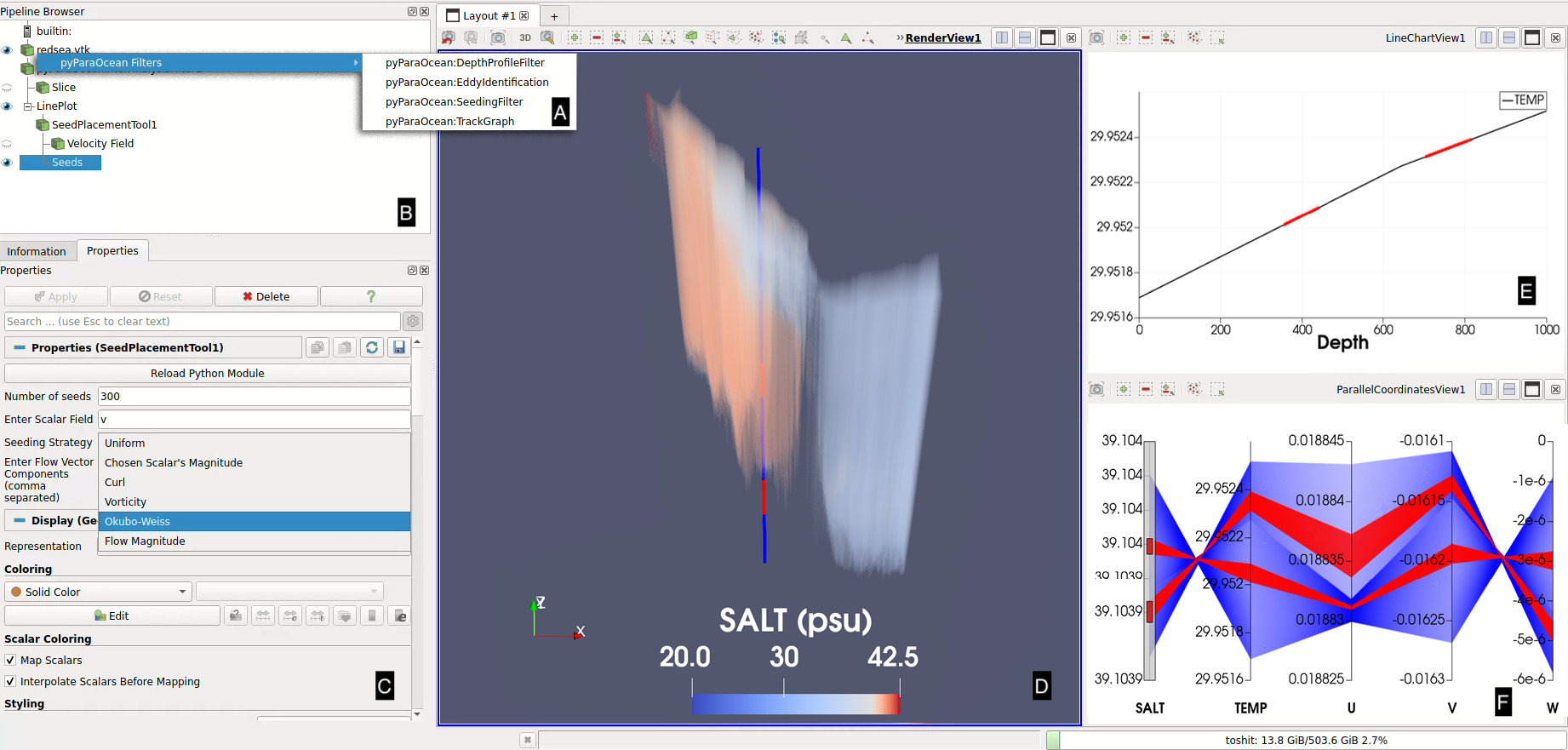}
    \includegraphics[width=.24\linewidth]{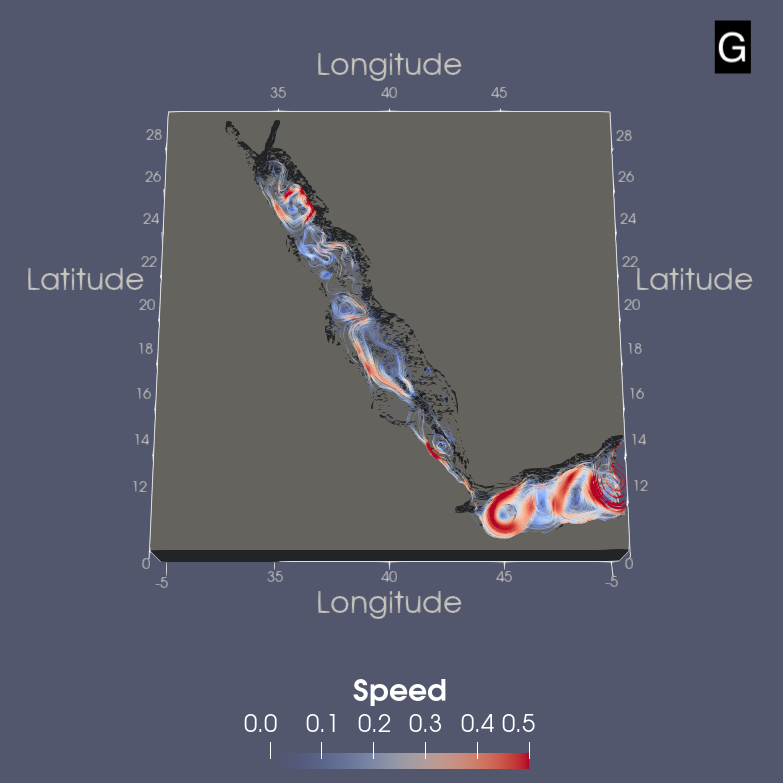}
    \includegraphics[width=.24\linewidth]{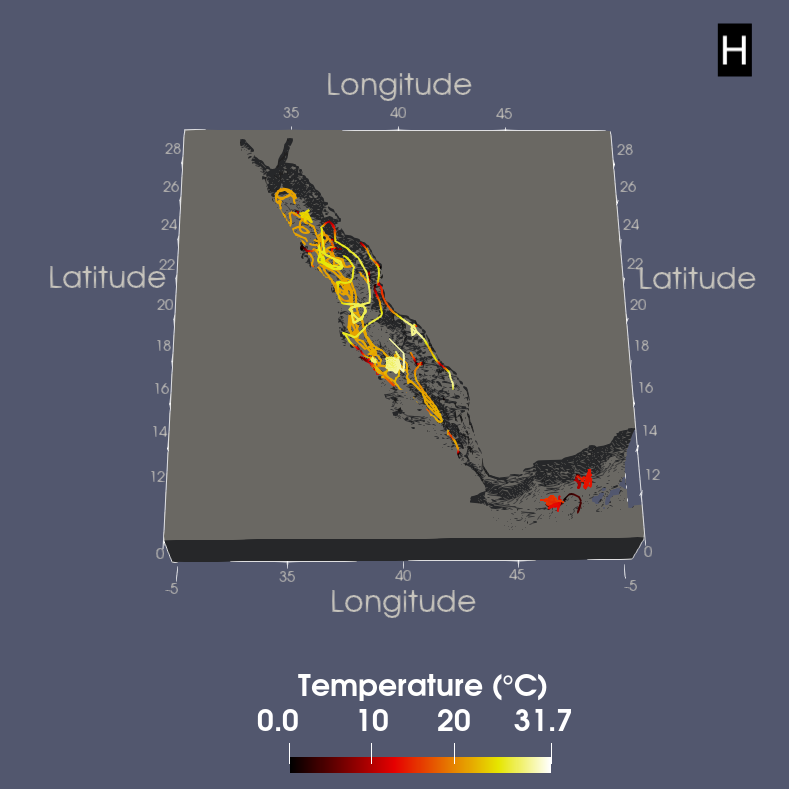}
    \includegraphics[width=.24\linewidth]{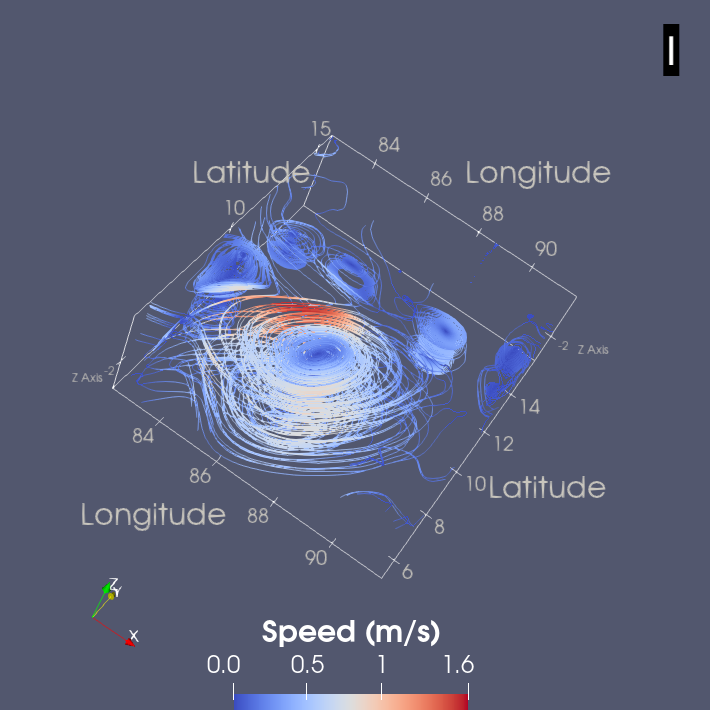}
    \includegraphics[width=.24\linewidth]{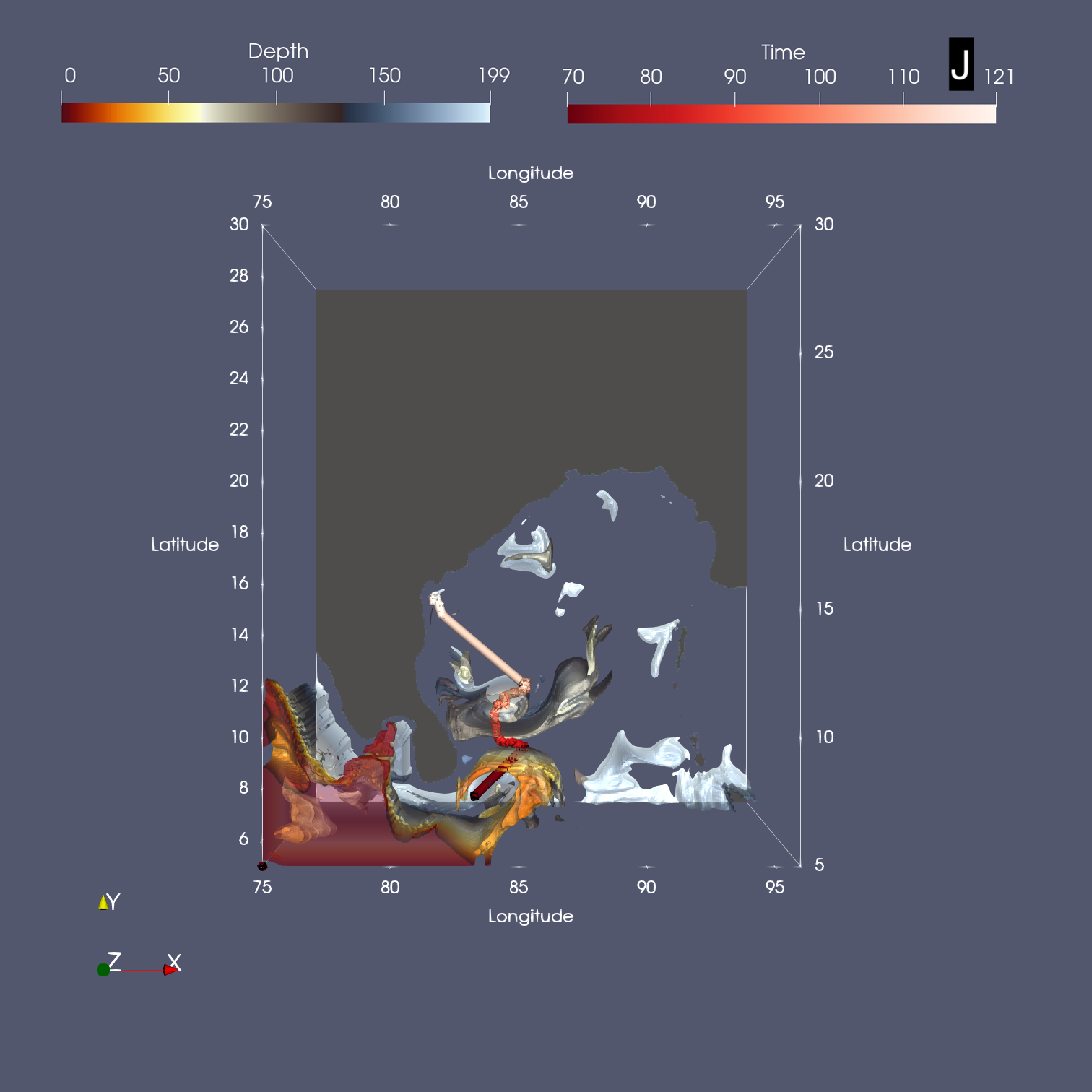}
    \caption{pyParaOcean functionality and user interface. (A)~All pyParaOcean modules are implemented as Paraview filters. (B)~Paraview pipeline browser shows the different datasets under study and the filters applied on them. (C)~The seeding filter from pyParaOcean provides multiple options for tracing field lines. The figure illustrates the usage of various filters showcasing (D)~salinity visualization in the Red Sea using volume rendering, (G)~flow visualization in the Red Sea with streamlines, (H)~interactively seeded pathlines in the Red Sea, (I)~eddy detection and visualization in the Bay of Bengal, and (J)~tracking high salinity water movement towards the east coast of India with a surface front track.}
    \label{fig:pyparaocean}
\end{figure*}
pyParaOcean is a plugin developed on top of Paraview to cater to the visualization needs of oceanographers, see Figure~\ref{fig:architecture}. Paraview~\cite{AhrensGeveciParaview2005} is an open source visualization software. The visualization pipeline in Paraview is a data flow network of executable modules. A module is a functional unit in Paraview with zero or more input ports and zero or more output ports. A module in Paraview can perform one of three tasks: produce data~(source), handle and process incoming data~(filter), or render/produce image~(sink). Paraview is a general purpose visualization tool and hence includes a large number of readers, data sources, and filters. The sheer number of available filters tends to get overwhelming and arduous to navigate, particularly for an application domain expert. Paraview also provides a mechanism for including new modules via plugins in scenarios where the available collection of modules does not meet the needs of the user. 

Figure~\ref{fig:architecture} shows an overview of the pyParaOcean architecture. The plugin is a collection of filters that provide visualization tools for interactive analysis of 3D ocean data and specialized features such as eddy computation and visualization, high salinity water visualization, and tracking of high salinity water masses. Ocean data is typically available as samples on a rectilinear grid in the NetCDF format. The plugin uses built-in VTK and Paraview libraries when appropriate, which are well maintained and supported, and hence benefits from the availability of support from the user community. The plugin can be loaded in a few simple steps.

Paraview is designed as a three-tier client-server architecture, consisting of a data server, a render server, and the client. The data server is responsible for all data related tasks such as reading, writing, and filtering. The render server is responsible for rendering and the interaction and exploration is performed at the client. The client manages the creation, execution, and destruction of objects on the servers, but it does not contain the data. This architecture is useful for running applications in a parallel environment. The data and render server can be run on a headless server or a supercomputer while the device with the end user acts as a client. When executed on the client without any connection to a remote server, Paraview seamlessly connects to a built-in server and provides all of its functionality. pyParaOcean is designed to leverage the parallelization capabilities of Paraview, to scale with larger data sizes. pyParaOcean aims to provide a set of filters specific to the needs of an oceanographer that is extensive, accessible, and easy-to-use.

\section{pyParaOcean: Functionalities}\label{sec:pyparaocean-function}
We now list and describe the various functions that are implemented in the pyParaOcean plugin and made available as Paraview filters. Figure~\ref{fig:pyparaocean} and the video in the supplementary material show the different filters in pyParaOcean and the user interface.

\subsection{Isovolume visualization}\label{subsec:VR}
Volume rendering is a natural choice for visualizing the 3D scalar fields in the ocean data because it provides a quick overview of the distribution (Figure~\ref{fig:pyparaocean}(D)). Animation with a fixed transfer function allows visualization of the scalar field over time. The volume rendering filter within Paraview can be tuned to visualize subvolumes of interest by choosing a interval within the range of the scalar field. Specifically, an isovolume containing the mean salinity/temperature value within the spatial region of interest or an isovolume that captures high salinity water provide a good overview of the 3D field.

\subsection{Seed placement and field lines}\label{subsec:SPT}
Field lines, including streamlines and pathlines, provide a good overview of a 3D vector field. pyParaOcean provides a filter that implements multiple seeding strategies for initiating the computation of streamlines and pathlines, and allows the user to choose one. Seeds generated using this filter are fed as input to the custom source streamline integrator or particle tracer in Paraview (Figure~\ref{fig:pyparaocean}(G,H)). 
   
Streamlines are a set of integral curves tangent to the velocity at every point in space. They portray instantaneous lines of flow that characterize important oceanographic phenomena such as eddies, currents, and filaments. Pathlines are tangential to the velocity as it evolves over time. A pathline describes the path a massless virtual particle would take beginning from the seed positioned at a particular timestep. Pathlines are useful for understanding transport, such as salinity advection and debris collection. They are more compute-intensive than streamlines.

The seeding filter controls the number of seeds and how the domain is sampled for seed placement, see Figure~\ref{fig:pyparaocean}~(C)). Sampling can be (a)~uniform, (b)~weighted by flow speed, curl, vorticity, or the Okubo-Weiss criterion~\cite{okubo1970horizontal}, or (c)~weighted by user-defined scalar fields that are computed earlier in the pipeline. A user may tune the line integration parameters and the sampling options to reduce visual clutter, focus the computation to regions of interest, maximize coverage of the domain, and highlight interesting flow features. For example, rendering wispy streamlines in regions of high vorticity produces closed loops around eddies with some temporal coherence across frames (Figure \ref{fig:cs1}). 

Additionally, each component of the vector field to which the seeding filter is applied can be specified as a separate scalar field. This makes it easier to perform other downstream operations like ignoring the vertical velocity component or adjusting the scale along each axis.

\subsection{Interactive particle paths}
This filter facilitates heat and mass transport queries on ocean data with an interactive seeding extension to the particle tracer from Paraview. It displays a linked parallel coordinates plot, where the user can brush to select ranges of scalars like temperature and salinity and hence restrict the seeding to isovolumes. Points sampled from these subvolumes serve as  seeds for pathline computation (Figure~\ref{fig:pyparaocean}(H)).

\subsection{Depth profile view} \label{subsec:depthprofile}
This filter enables the user to inspect a vertical column of the ocean, specified by a longitude and latitude pair.  It drops a ``needle'' into the ocean and samples points along this line at different values of depth (Figure~\ref{fig:pyparaocean}(D)). It displays a linked parallel coordinates plot that gives a depth profile of all scalars sampled along the vertical column. A line plot view of the chosen scalar against depth (Figure~\ref{fig:pyparaocean}(E)) is displayed. Optionally, the scalar field mapped to a vertical slice at the chosen longitude is shown in the volume render window. The user can select and highlight a subset of the points in the vertical column from the parallel coordinates plot (Figure~\ref{fig:pyparaocean}(F)), and track them across time in all views. This is useful for studying vertical mass transport, especially upwelling or downwelling via Ekman transport~\cite{sarmiento2013ocean} in eddy centres, and to study the depression of isotherms indicating redistribution of heat \cite{kumar2007eddy}. Studying these changes is important to understand marine life as they drive phytoplankton bloom and nutrient transport.

\begin{figure}
    \centering
    \subcaptionbox{\label{subfig:minima_unfiltered}}{\includegraphics[width=.48\linewidth]{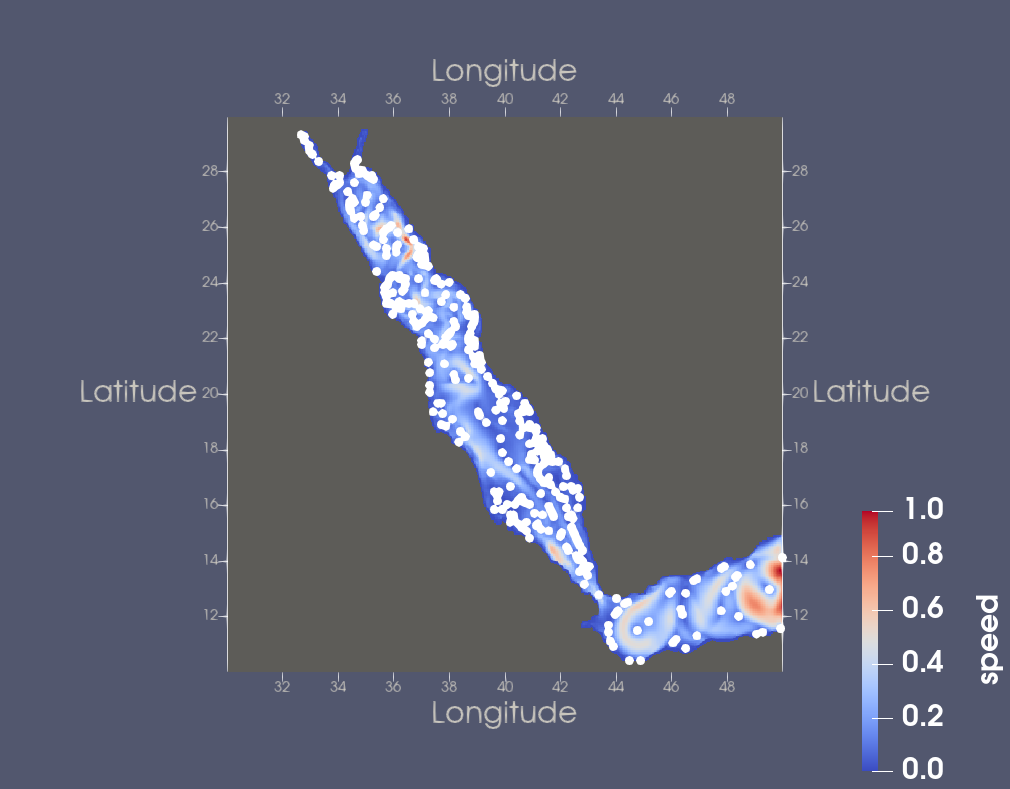}}
    \subcaptionbox{\label{subfig:minima_filtered}}{\includegraphics[width=.48\linewidth]{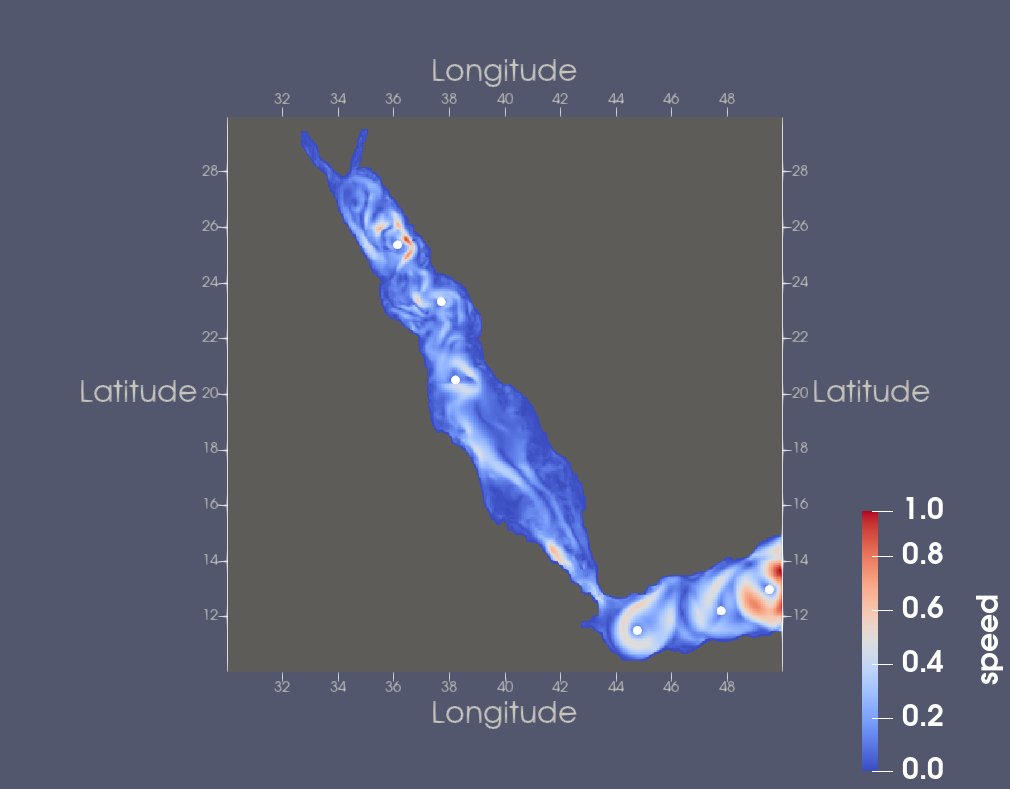}}
    \caption{Efficacy of the method used for eddy centre detection. (a)~Local minima (white spheres) of the velocity magnitude are potential eddy centres. (b)~Topological simplification removes noise and insignificant minima, and identifies vortex cores of mesoscale eddies in the Red Sea and Gulf of Aden.}
    \label{fig:eddy-centres-redsea}
\end{figure}
\begin{figure*}[h]
    \centering
    \includegraphics[width=.22\linewidth]{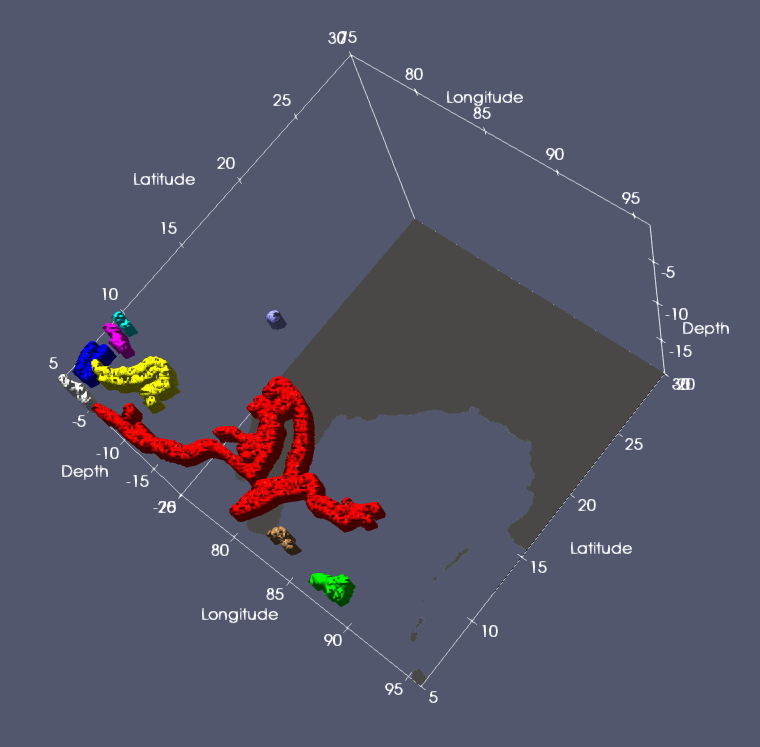}
    \includegraphics[width=.22\linewidth]{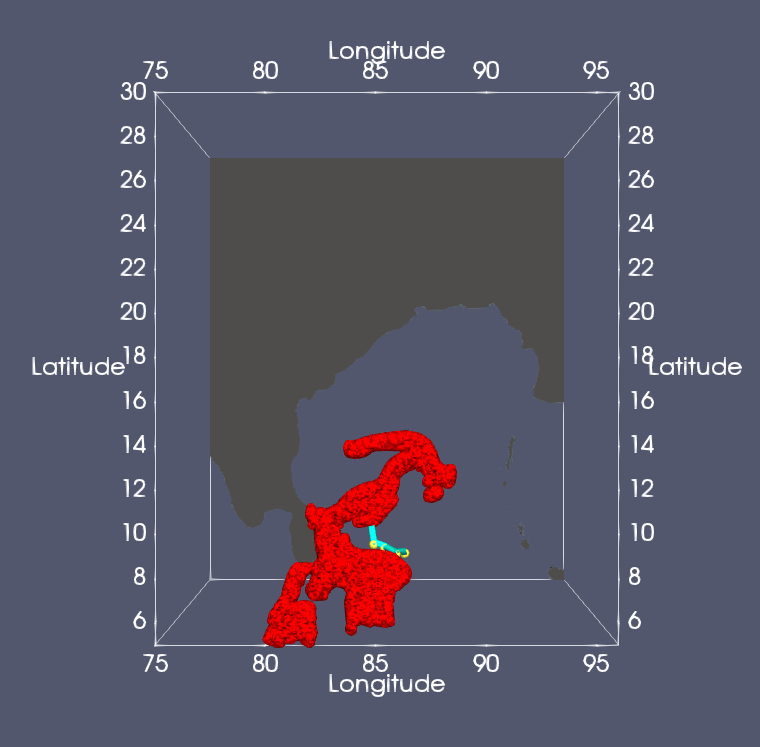}
    \includegraphics[width=.22\linewidth]{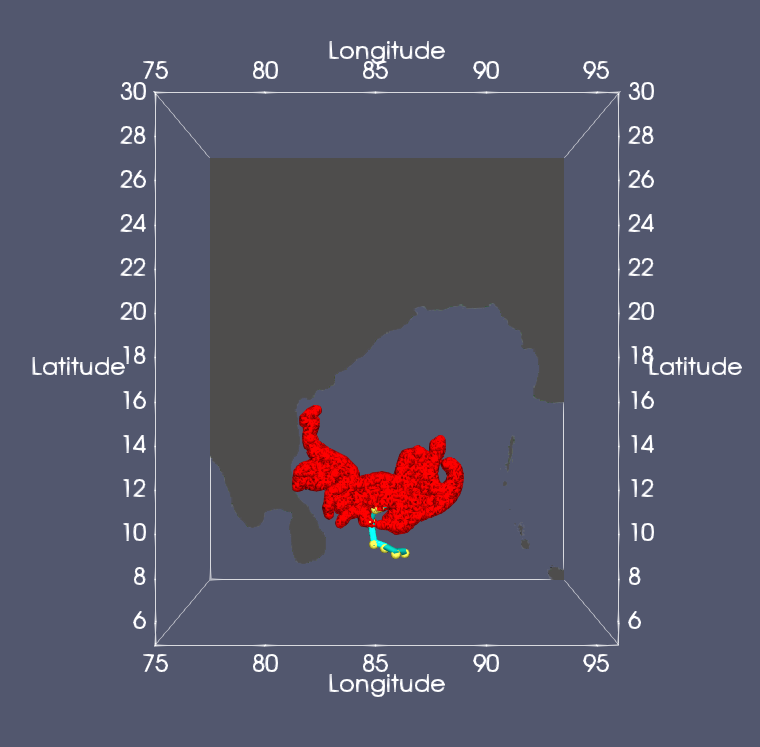}
    \includegraphics[width=.22\linewidth]{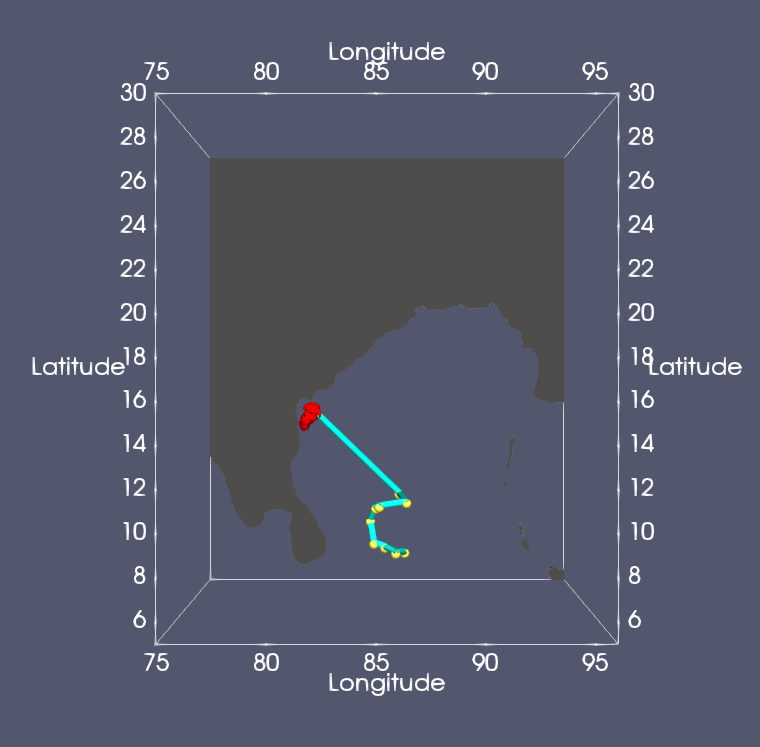}
    \caption{Visualizing movement of high salinity water via computation and tracking of surface fronts of high salinity isovolumes. (left)~Surface fronts computed at one time step. (middle, right)~One of the components of the surface front moves towards the east coast of India, near Visakhapatnam. The evolution of this surface front component is computed and visualized as a track.}
    \label{fig:hsc-front-track}
\end{figure*}
\subsection{Eddy identification and visualization}
Several algorithms have been devised for reliable and automatic identification of eddies~\cite{afzal2019state}. McWilliam~\cite{mcwilliams1990vortices} developed a 2D method  using vorticity $\omega$ as a physical parameter, whose local minima and maxima locate the centres of potential eddies and the values of vorticity in the neighborhood relative to the centre helps determine the eddy boundary. 
%The boundary of the eddy is determined by the values of vorticity around the centre $\omega / \omega_{cen} < 0.2$, where $\omega_{cen}$ is the magnitude of vorticity at the centre and making sure that this boundary is symmetric around the centre.
Okubo~\cite{okubo1970horizontal} uses a special Okubo-Weiss parameter based on shear and strain deformation and the vertical component of vorticity to measure rotation and hence identify potential eddies. A circularity criterion may be applied after the Okubo-Weiss criterion to improve the results~\cite{williams2011visualization}. Sea surface height and velocity profile have also been used for eddy detection~\cite{matsuoka2016new}. The winding angle criterion, together with a streamline clustering,  helps identify eddies in 3D~\cite{friederici2021winding}.

The eddy identification filter in pyParaOcean focuses on mesoscale eddies~\cite{amores2017coherent} . It uses only the velocity field in individual time steps and does not compute any derived fields. This 3D detection scheme can be applied in parallel across timesteps and across depth slices since the vertical velocity is not used.

Flow speed of swirling fluid decreases radially inwards towards the centre of rotation. The filter inspects local minima of the flow speed to identify potential eddy centres. Vertical velocity is ignored to discount the motion of upwelling or downwelling in vortex cores, thus enhancing the corresponding flow minima. Noise and less significant minima are removed by applying topological simplification directed by the notion of persistence~\cite{tierny2017topology}. Next, the method employs an approximation of the winding angle criterion~\cite{friederici2021winding} by checking if the streamline crosses into all four quadrants of an XY plane centred at the minimum~\cite{daguo04cross}. This method is more effective in regions with relatively stationary eddy centres like the Red Sea and Gulf of Aden. Figure~\ref{fig:eddy-centres-redsea} shows the set of potential eddy centres identified in the Red Sea using this filter. 

Streamlines seeded near the core of an eddy form spirals or closed loops. The boundary of an eddy is determined using a binary search along the radial axes. The search helps locate the seed that is furthest from the eddy centre but results in a spiral or nearly closed loop streamline. The filter displays all streamlines originating near the detected vortex core and hence presents a 3D profile of the eddy (Figure \ref{fig:pyparaocean}(I)). It may be extended to support other eddy detection methods~\cite{matsuoka2016new,friederici2021winding} that may be selected via the interface.

\subsection{Surface front tracking and salinity visualization}
Oceanographers are often interested in water parcels that transport mass or heat. These are moving volumes of water with distinctive temperature or salinity characteristics. The surface front tracking filter computes connected components of the boundary of a scalar field isovolume (called the surface front), tracks them over time, and generates a track graph that summarizes movement of all surface fronts. A subset of tracks extracted from this graph are rendered for visual analysis. Surface fronts have been shown to be a good representative of high salinity water masses~\cite{SINGH2022104993}. It has been used to trace the path of the high salinity core (HSC) entering into the Bay of Bengal from the Arabian Sea (Figure~\ref{fig:pyparaocean}(J)).

\begin{figure}
    \centering
    \includegraphics[width=0.7\linewidth]{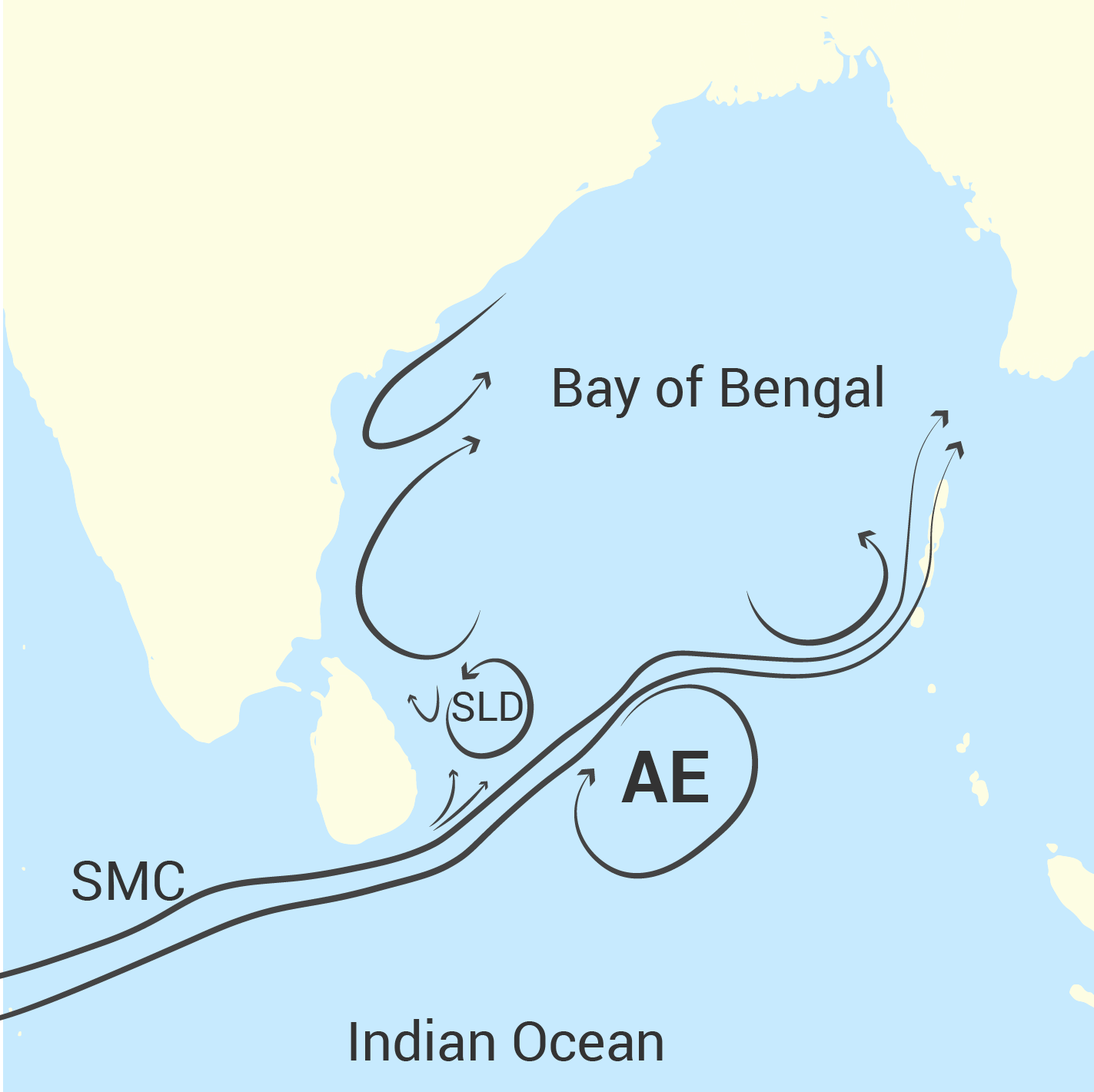}
    \caption{Currents and eddies in the Bay of Bengal during the monsoon season, including the Summer Monsoon Current (SMC), the Sri Lanka Dome (SLD), and an anticyclonic eddy (AE).}
    \label{fig:bobSketch}
\end{figure}

\begin{figure*}[!h]
\centering
    \subcaptionbox{\label{subfig:cs3:1}}{\includegraphics[width=.33\linewidth]{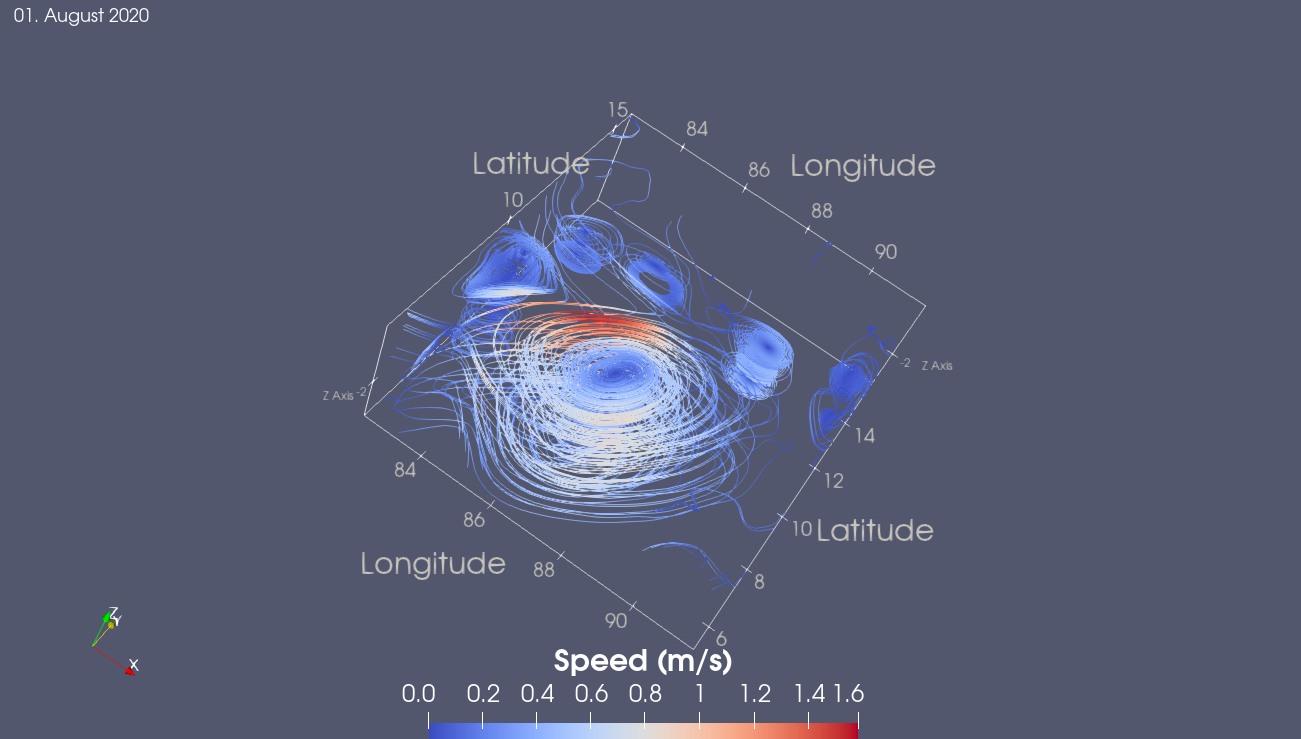}}
    \subcaptionbox{\label{subfig:cs3:2}}{\includegraphics[width=.33\linewidth]{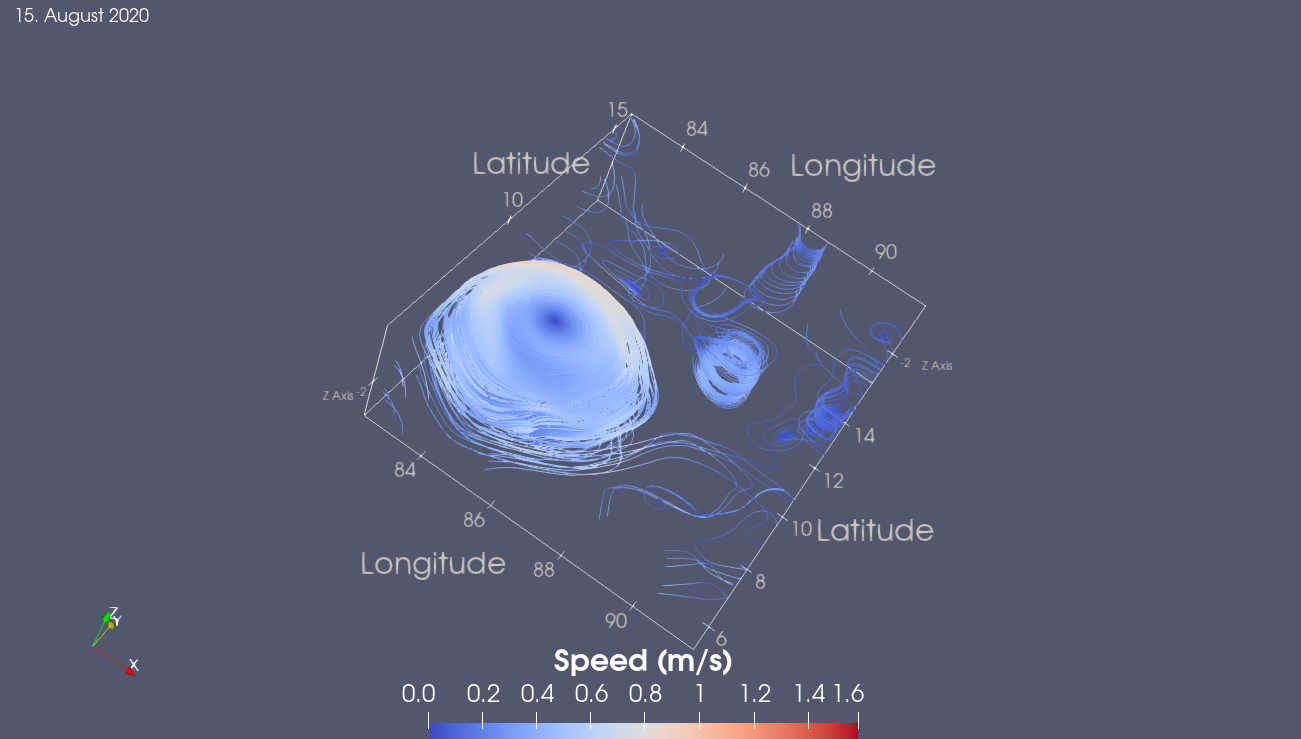}}
    \subcaptionbox{\label{subfig:cs3:3}}{\includegraphics[width=.33\linewidth]{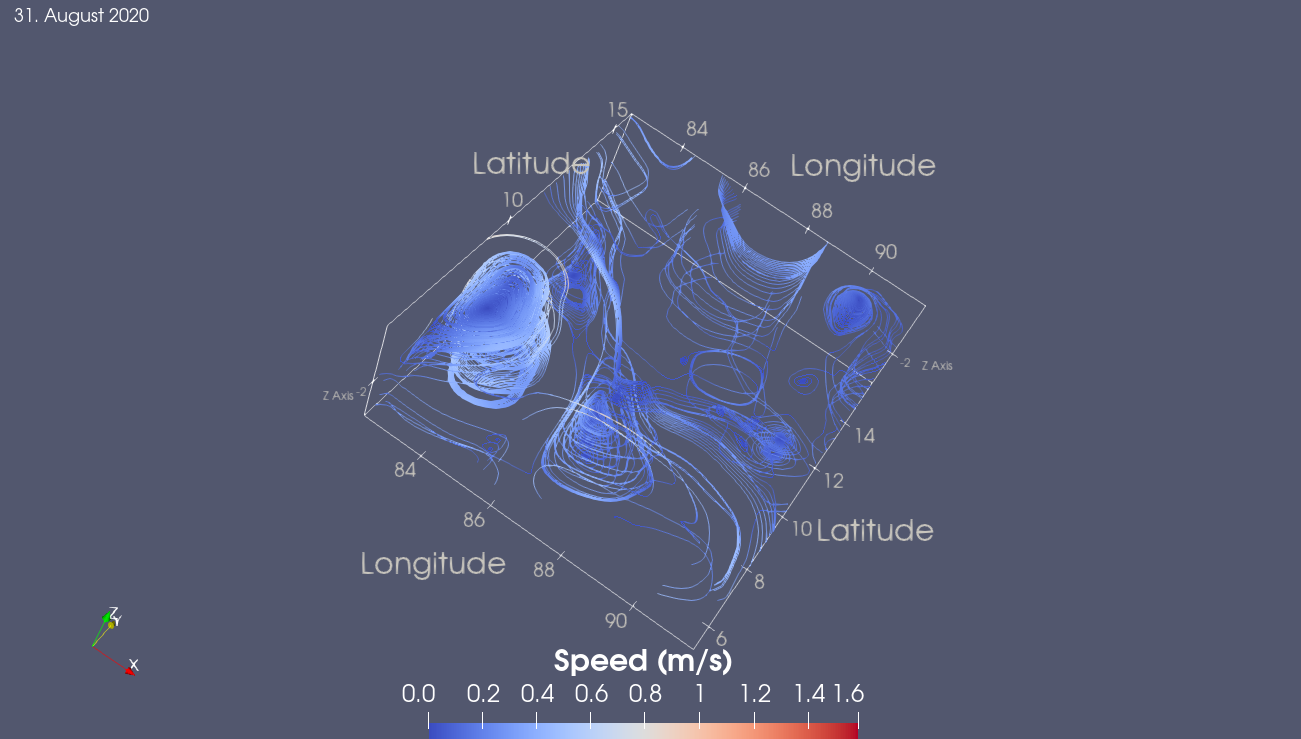}}
    \caption{Dissipation of a large anticylonic eddy in the Bay of Bengal through August 2020. Streamlines are seeded near detected vortex cores to show evolution of eddy profiles in 3D.}
   \label{fig:cs3}
\end{figure*}
\begin{figure*}[!h]
\centering
    \subcaptionbox{\label{subfig:cs1:1}}{\includegraphics[width=.33\linewidth]{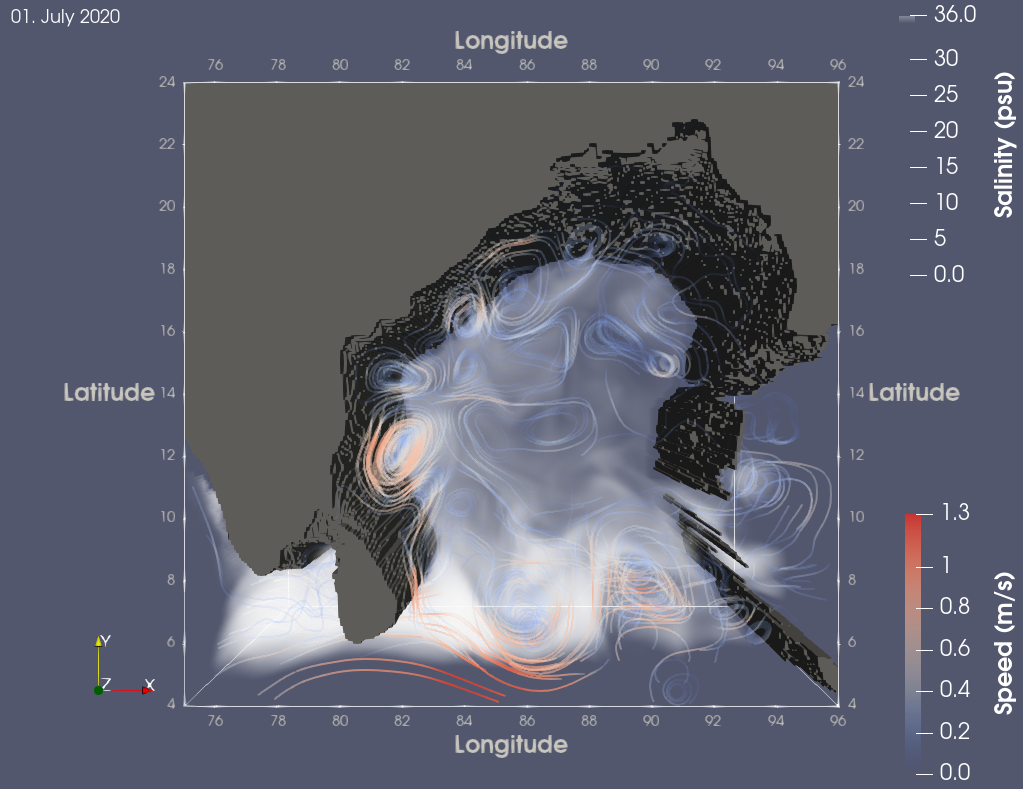}}
    \subcaptionbox{\label{subfig:cs1:2}}{\includegraphics[width=.33\linewidth]{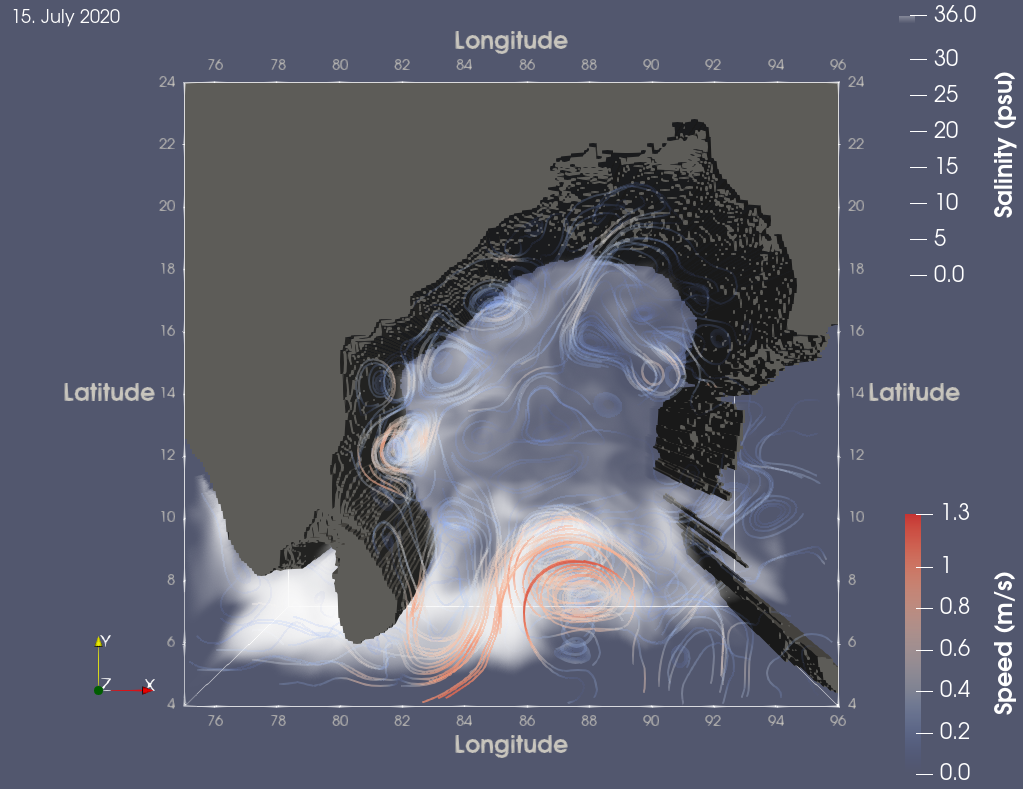}}    
    \subcaptionbox{\label{subfig:cs1:3}}{\includegraphics[width=.33\linewidth]{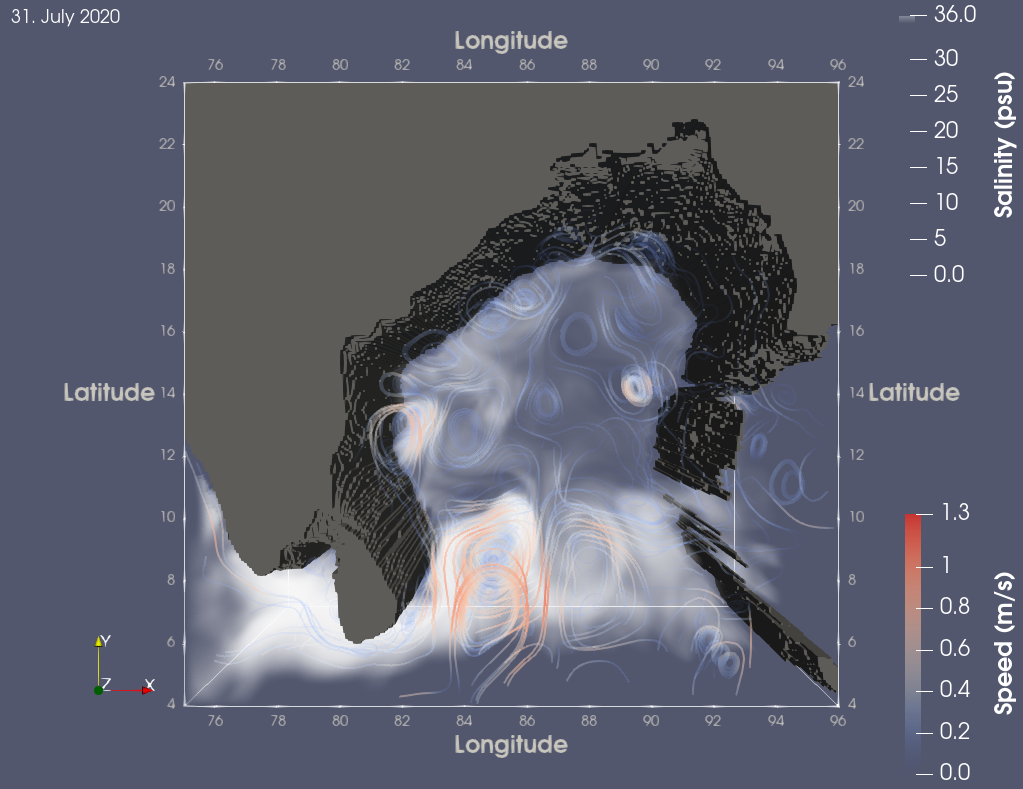}}

    \caption{The Bay of Bengal between July 1, 2020 and July 31, 2020. Visualization of the flow using streamlines with uniform seeding and ($\geq 35$~psu) salinity isovolume rendering. (a)~July 1, 2020: The AE can be seen forming around 8\textdegree N and 90\textdegree E with the SMC streamlines visible from 78\textdegree E to about 86\textdegree E. (b)~July 15, 2020: The AE, 8\textdegree N and 87\textdegree E, has matured into a circular shape and moves westward towards Sri Lanka. The ($\geq 35$~psu) isovolume shows recirculation of high salinity waters into they Bay by AE. (c)~July 31, 2020: The AE, 7\textdegree N and 84\textdegree E, reaches the eastern coast of Sri Lanka where it will start to dissipate.}
    \label{fig:cs1}
\end{figure*}
\begin{figure*}[!h]
\centering
    \subcaptionbox{\label{subfig:cs2:1}}{\includegraphics[width=.45\linewidth]{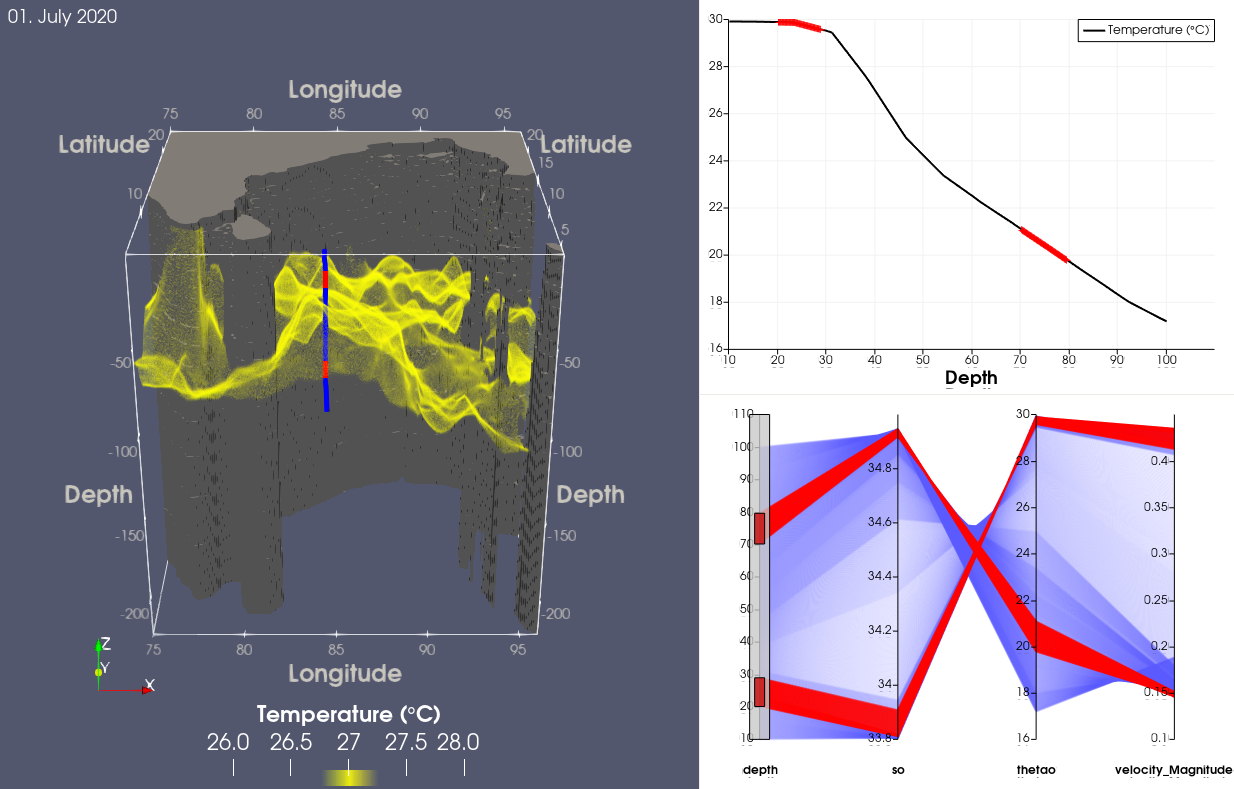}}
    \subcaptionbox{\label{subfig:cs2:2}}{\includegraphics[width=.45\linewidth]{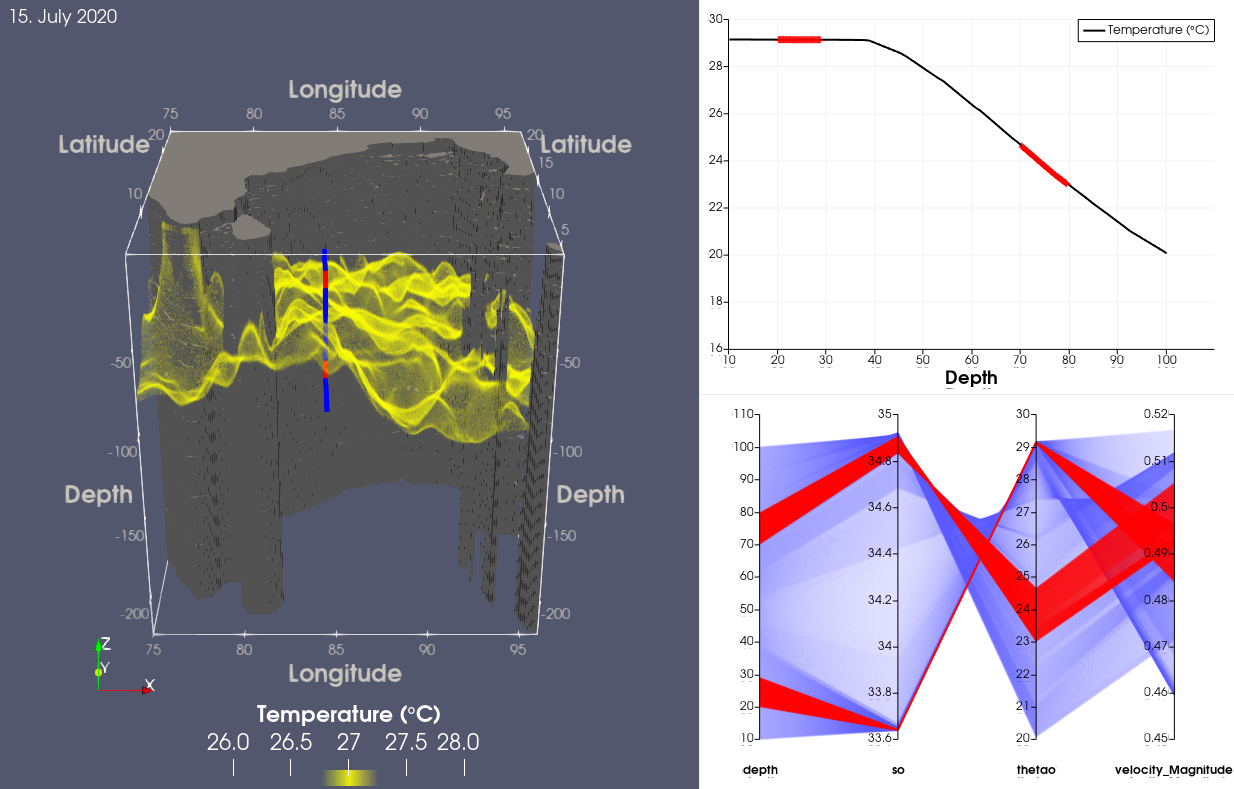}}
    \subcaptionbox{\label{subfig:cs2:3}}{\includegraphics[width=.45\linewidth]{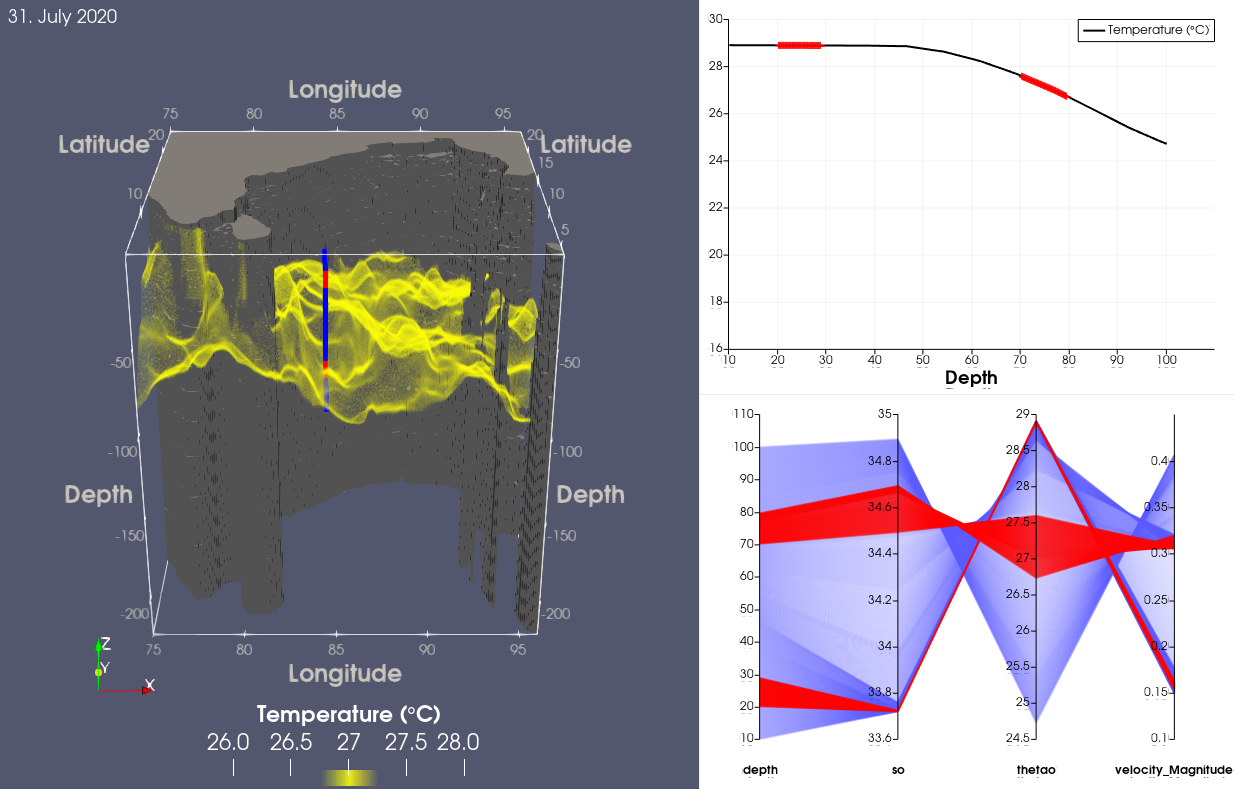}}

    \caption{The depression of the 27$^{\circ}$ isotherm (yellow) by the anticylonic eddy in the Bay of Bengal. A needle is dropped at 7$^{\circ}$N, 84$^{\circ}$E and the depth profile shows the temperature drop. The interactive parallel coordinates plot is used to brush-select 10~m intervals at depths of 25~m and 85~m. (a)~July 1, 2020: The downwelling of the AE can be seen around 8$^{\circ}$N and 90$^{\circ}$E, at the depth of 100~m. As it forms, the AE pushes the 27$^{\circ}$ isotherm down. (b)~July 15, 2020: The AE, 8$^{\circ}$N and 87$^{\circ}$E, can be seen moving east with the depression of the isotherm and the depth profile of temperature begins to flatten near 29$^{\circ}$C as the eddy moves closer to the needle. (c)~July 31, 2020: The AE centre, 7$^{\circ}$N and 85$^{\circ}$E, is very close to the needle and the depression in the isotherm has moved all the way to near the east coast of Sri Lanka.}
    \label{fig:cs2}
\end{figure*}
%

%

%-------------------------------------------------------------------------
\section{Case study: Bay of Bengal}\label{sec:casestudy}
%\subsection{Bay of Bengal - Anticyclonic Eddy}
The Summer Monsoon Current (SMC) is a prominent feature of Indian ocean circulation and the SMC flows around Sri Lanka to flow into the Bay of Bengal. We use pyParaOcean to study different phenomena in the Bay of Bengal, particularly during the monsoon.

\myparagraph{Eddies.} 
Figure~\ref{fig:bobSketch} is a rough schematic of the major currents and eddies in the Bay during the monsoon season. A large anticyclonic eddy (AE)  located to the right of the SMC and a cyclonic eddy known as the the Sri Lanka Dome (SLD) to its left~\cite{vinayachandran1998monsoon} are regular features in this region during summer. The AE has a diameter of about 500~km, located to the southeast off the coast of Sri Lanka, and is characterized by intense downwelling inside owing to its anticyclonic circulation. \cite{vinayachandran1998monsoon} proposed that the AE is formed by the interaction of the SMC and the incoming Rossby waves from Sumatra. The timeline of appearance and disappearance of the AE was documented in later work~\cite{vinayachandran2004biological}. The AE starts forming in June, develops into its circular shape in July, and weakens in August, as shown in Figure~\ref{fig:cs3} and the accompanying video.

\myparagraph{Salinity transport.} 
pyParaOcean serves as an efficient tool to analyze the effects of AE on the Bay of Bengal. Streamlines and pathlines offer visualization of circulation associated with the AE and its movement in the ocean. The field lines may be overlaid on a volume rendering of a scalar to visualize the transport caused by the eddy. Figure~\ref{fig:cs1} and the accompanying video show the streamlines overlaid on a salinity volume rendering at different time steps to show the role of the AE in transport of salt. The movement of high salinity water from the Arabian sea by the SMC into the Bay of Bengal and its recirculation by the AE is well captured in this representation. Tracking surface fronts of high salinity water and highlighting the long-lived tracks helps capture an overview of significant salinity movement in the region. We observe a track that moves towards the coast of India, see Figure~\ref{fig:hsc-front-track}.

\myparagraph{Downwelling.} 
Figure~\ref{fig:cs2} and the accompanying video show the use of the depth profile filter to visualize the depression of the 27$^{\circ}$ isotherm by the AE. The anticyclonic nature of the eddy causes a downwelling inside the eddy and pushes the relatively warmer water downward. The parallel coordinates view shows changes in temperature, salinity, and speed in the water column caused by the arrival of the eddy at the point of interest.

\myparagraph{Experience and performance.} 
This case study was conducted in collaboration with an oceanographer coauthor. Several observations on phenomena such as the SLD and the movement of high salinity water could be made using pyParaOcean. While our oceanographer collaborators typically use tools such as pyFerret for 2D analysis, they found the capability of pyParaOcean to be very useful. After this initial satisfying experience, we plan to work together on the study of higher resolution model output using pyParaOcean. The surface front tracking and eddy detection filters take a few minutes, while all other filters take 1-2 seconds or less. All the experiments were run on a workstation with an 8 core AMD EPYC 7262 @ 3.2 GHz CPU with 512 GB main memory and  NVIDIA RTX A4000 (16 GB) GPU. The surface front computation is parallelized using the python multiprocessing library but there is scope for further improvement in runtime. The eddy detection and visualization filter can also be optimized by parallelizing some of the computation. We plan to take this up in the future.

%-------------------------------------------------------------------------
\section{Conclusions}\label{sec:conclusions}
This paper described pyParaOcean, a system for visual exploration and analysis of 3D time-varying ocean data. pyParaOcean is designed as a plugin for Paraview and provides visualization and analysis functionalities that are specific to oceanography. Its integration into Paraview enables efficient computation and optional support for a larger set of general purpose filters for visual analysis. A case study on the Bay of Bengal, conducted together with an oceanographer collaborator, illustrates the use of pyParaOcean to explore eddies, ocean currents, and salinity movement in the Bay of Bengal. In future work, we will include support for additional filters for computing statistics, temporal data analysis, and correlation studies. We also plan to improve the runtime performance and conduct scalability tests for higher resolution ocean models before releasing the plugin as an open source software.    

%-------------------------------------------------------------------------

\section*{Acknowledgments}
This research was funded by a grant from SERB, Govt. of India (CRG/2021/005278), partial support from National Supercomputing Mission, DST, the J. C. Bose Fellowship awarded by the SERB, DST, Govt. of India, the Dr. Ram Kumar IISc Distinguished Visiting Chair Professorship in EECS, and a scholarship from MoE, Govt. of India. Part of this work was carried out towards partial fulfilment of a thesis requirement at BITS Pilani.

\bibliographystyle{eg-alpha-doi}

\bibliography{references}

%---------------------------------------------------------------------

\end{document}